\documentclass{article}
\usepackage{todonotes}
\usepackage{ijcai19}
\usepackage{times}
\usepackage{soul}
\usepackage{url}
\usepackage[utf8]{inputenc}
\usepackage{booktabs}
\usepackage[small]{caption}
\usepackage{verbatim}
\usepackage{natbib}
\usepackage{setspace}
\usepackage{fancyhdr}
\usepackage{subcaption}
\usepackage{adjustbox}
\usepackage{tabularx}
\usepackage[noend]{algpseudocode}
\usepackage{multirow}
\usepackage{graphics}
\usepackage{graphicx}
\usepackage{color}
\usepackage{amsmath}
\usepackage{algorithm}
\usepackage{amssymb}
\usepackage{graphicx,wrapfig,lipsum}
\usepackage[hidelinks=true]{hyperref}

\newtheorem{theorem}{Theorem}[section]

\newtheorem{hypothesis}{Hypothesis}

\usepackage{xcolor,colortbl}

\definecolor{Gray}{gray}{0.85}
\definecolor{LightCyan}{rgb}{0.88,1,1}

\newcolumntype{a}{>{\columncolor{Gray}}c}
\newcolumntype{b}{>{\columncolor{LightCyan}}c}

\newcommand{\system}{DeFrauder}
\newcommand{\groupdetector}{\texttt{GroupDetector}}
\newcommand{\reviewervec}{\texttt{Reviewer2Vec}}

\title{Spotting Collective Behaviour of Online Frauds in Customer Reviews}

\author{
Sarthika Dhawan$^1$\and
Siva Charan Reddy Gangireddy$^1$\and \\
Shiv Kumar$^2$\And
Tanmoy Chakraborty$^1$\\
\affiliations
$^1$Indraprastha Institute of Information Technology Delhi (IIITD), India\\
$^2$Netaji Subhas University of Technology (NSUT), Delhi, India\\
\emails
\{sarthika15170, sivag\}@iiitd.ac.in,
shivk.it.16@nsit.net.in,
tanmoy@iiitd.ac.in
}

\begin{document}

\maketitle
\begin{abstract}
Online reviews play a crucial role in deciding the quality before purchasing any product.
Unfortunately, spammers often take advantage of online review forums by writing fraud reviews to promote/demote certain products. It may turn out to be more detrimental when such spammers collude and collectively inject spam reviews as they can take complete control of users' sentiment due to the volume of fraud reviews they inject.  Group spam detection is thus more challenging than individual-level fraud detection due to unclear definition of a group, variation of inter-group  dynamics, scarcity of labeled group-level spam data, etc. Here, we propose \system, an unsupervised method to detect online fraud reviewer groups. It first detects candidate fraud groups by leveraging the underlying product review graph and incorporating several behavioral signals which model multi-faceted collaboration among reviewers. It then maps reviewers into an embedding space and assigns a spam score to each group such that groups comprising spammers with highly similar behavioral traits achieve high spam score. While comparing with five baselines on four real-world datasets (two of them were curated by us), \system\ shows superior performance by outperforming the best baseline with 17.11\% higher NDCG@50 (on average) across datasets.      
\end{abstract}

\section{Introduction}

Nowadays, online reviews are becoming highly important for customers to take any purchase-related decisions. Driven by the immense financial profits from product reviews, several blackmarket syndicates facilitate to post deceptive reviews to promote/demote certain products.  Groups of such fraud reviewers are hired to take complete control of the sentiment about products. Collective behaviour is therefore more subtle than individual behaviour. At individual level, activities might be normal; however, at the group level they might be substantively different from the normal behavior. Moreover, it is not possible to understand the actual dynamics of a group by aggregating the behaviour of its members due to the complicated, multi-faceted, and evolving nature of inter-personal dynamics. Spammers in such collusive groups also adopt intelligent strategies (such as paraphrasing reviews of each other, employing a subset of their resources to one product, etc.)  to evade detection. 

Previous studies mostly focused on individual-level fraud detection \citep{Lim_CIKM10,Fei_AAAI,Akoglu_ICWSM}. Few other studies which realized the detrimental effect of such collective activities detected groups simply based on Frequent Itemset Mining (FIM)  \citep{Arjun_www,xu2013uncovering,allahbakhsh2013collusion}. They thus focused more on ranking fraud groups, paying less attention to judge the quality of the detected groups.

\citep{Song} pointed out several limitations of  FIM for group detection -- high computational complexity at low minimum support, absence of temporal information, unable to capture overlapping groups, prone to detect small and tighter groups, etc.

\begin{figure*}[!t]
    \centering
    \scalebox{0.6}{
    \includegraphics{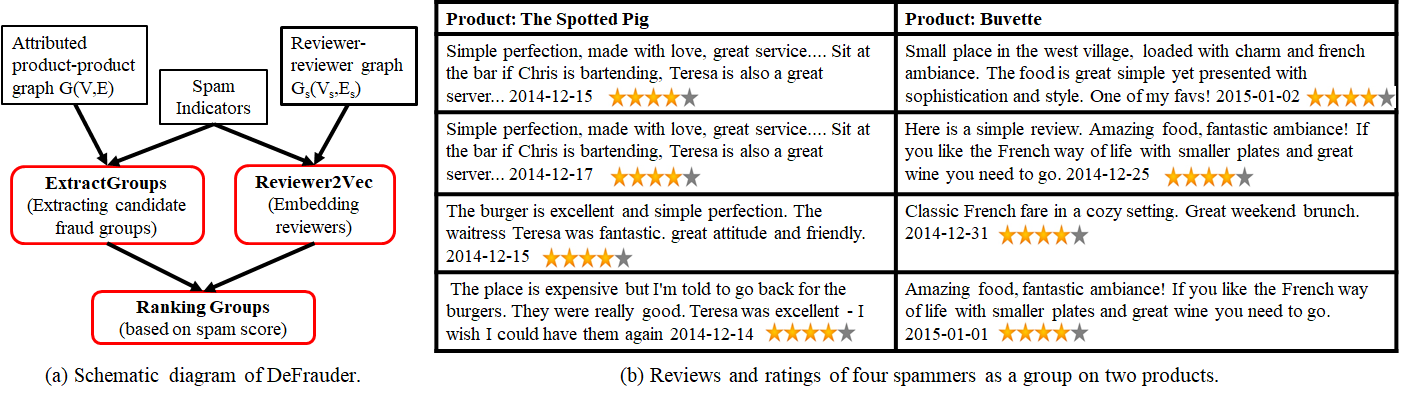}}
    \caption{(a) Workflow of \system. (b) Coherent review patterns of a fraud reviewer group on two products. To evade detection, four reviewers in a group paraphrase reviews of each other keeping the underlying sentiment same. }
    \label{fig:example}
\end{figure*}

In this paper, we propose \system\footnote{\system: {\bf De}tecting {\bf Fraud} R{\bf e}viewer G{\bf r}oups, Code is available in \citep{si}. },
a novel architecture for fraud reviewer group  detection. 
\system\ contributes equally to (i) the detection of potential fraud groups by incorporating several coherent behavioral signals of  reviewers, and (ii) the ranking of groups based on their degree of spamicity   by proposing a novel ranking strategy. Experiments on four real-world labeled datasets (two of them were prepared by us) show that \system\ significantly outperforms five baselines  -- it beats the best baseline by 11.35\% higher accuracy for detecting groups, and 17.11\% higher NDCG@50 for ranking groups (averaged over all datasets). 

In short, our contributions are fourfold: (i) two novel datasets, (ii) novel method for reviewer group detection, (iii) novel method for ranking groups, and (iv) a comprehensive evaluation to show the superiority of \system.

\section{Related Work}\label{sec:relatedwork}
Due to the abundance of literature \citep{kou2004survey} on online fraud detection, we restrict our discussion to {\em fraud user detection on product review sites}, which we deem as pertinent to this paper.  

In user-level fraud detection, notable studies include \citep{Lim_CIKM10} which proposed scoring methods to measure the degree of spamicity of a reviewer based on rating behaviors; \citep{Wang_ICDM} which developed a graph-based fraud reviewer detection model; \citep{Fei_AAAI} which exploited burstiness in reviews to detect fake reviewers. SpEagle \citep{Akoglu_ICWSM}  utilizes clues from all metadata as well as relational data and harnesses them collectively under a unified framework. ASM \citep{Arjun_ACM} facilitates modeling spamicity as latent factor and allows to exploit various behavioral footprints of reviewers. \citep{Guan_ACM} argued that the dynamic behavioral signals can be captured through a heterogeneous reviewer graph by considering reviewers, reviews and products together. FraudEagle \citep{Akoglu_ICWSM}  spots fraudsters and fake reviews simultaneously. \citep{dutta2018retweet, Chetan:2019} discussed about collusion in social media.

Few studies attempted to detect group-level fraud reviewer groups. GSRank \citep{Arjun_www} was the first method of this kind, which identifies  spam reviewer groups using FIM, and ranks groups based on the relationship among groups, products, and individual reviewers. \citep{Arjun_www} argued that due to the scarcity of data, unsupervised method should be used to tackle this problem.   Other studies largely focused on improving the ranking algorithm, ignoring the performance of the group detection method (they also used FIM for detecting groups)~\citep{xu2013uncovering,allahbakhsh2013collusion,mukherjee2013yelp,ott2011finding}.
\citep{Xu_IEEE} proposed an Expectation Maximization algorithm to compute the collusive score of each group detected using FIM. \citep{Song} argued that FIM tends to detect small-size and tighter groups. They proposed GSBP,  a divide-and-conquer  algorithm which emphasizes on the topological structure of the reviewer graph. GGSpam~\citep{Wang2018} directly generates   spammer groups consisting of group members, target products, and spam reviews.

\begin{table}[]
 \centering
\scalebox{1}{
      \begin{tabular}{l|c|c|c|c}
        \toprule  &GSBP&GGSpam&GSRank&\system \\ \midrule
        Temporal & 
        \checkmark&\checkmark&\checkmark&\checkmark \\  
        Content   & &&\checkmark&\checkmark \\  
        Graph   & \checkmark&\checkmark&&\checkmark \\ \bottomrule
          \end{tabular}}
          \caption{\system\ captures all three signals obtained from users attributes and underlying graph structure.}\label{tab:comparison}

\end{table}

We propose \system\ that leverages temporal pattern, content similarity and underlying graph structure to detect collusive fraud groups in review forums (see Table \ref{tab:comparison} for a comparison). Unlike others, \system\ contributes equally to both the components -- group detection and ranking.

\section{Proposed \system\ Framework}
In this section, we explain \system\, which is a three-stage algorithm -- detecting candidate fraud groups, measuring different fraud indicators of candidate groups, and ranking groups based on the group spam score.

\subsection{Group Fraud Indicators}\label{sec:indicator}
Here we present six fraud indicators which measure the spamicity of a group \citep{Arjun_www,Wang2018,Song} by taking into account linguistic, behavioral, structural and temporal signals. All indicators are normalized to $[0,1]$ -- larger value indicates more spamming activity. Let $R$ and $P$ be the entire set of reviewers and products. Let $R(g)$ be the set of reviewers in a group $g$, and  $P(g)$ be the set of target products reviewed by the reviewers in $R(g)$. Each reviewer $i$ reviewed a set of products $P_i$. 

To reduce the contingency of small groups, we use a penalty function $L(g)$ which is a logistic function \cite{Song}: 
$L(g)=1/(1+{e}^{\textbf{-}(|R(g)|+|P(g)|\textbf{-}3)})$. We subtract $3$ from the sum since we consider minimum number of reviewers and products within a group to be $2$ and $1$, respectively; therefore $L(g) \in  [0.5, 1)$ \cite{wang2018gslda}.

\noindent \paragraph{Review tightness(RT).} It is the ratio of the number of reviews  in $g$ (assuming each reviewer is allowed to write one review to each product) to the product of the size of reviewer set and product set in $g$.

\begin{equation}
    RT(g) =\frac{\sum_{i\in R(g)}|P_i|}{|R(g)||P(g)|}\cdot L(g)
\end{equation}
If significant proportion of people co-reviewed several products, then it may imply a spam group activity.

\noindent \paragraph{Neighbor tightness(NT).} It is defined as the average value of the Jaccard similarity (JS) of the product sets of each reviewer pair. 
\begin{equation}
NT(g)=\frac{\sum_{i,j \in R(g)} JS(P_i,P_j)}{\binom{|R(g)|}{2}}\cdot L(g)
\end{equation}
If the product sets are highly similar, both the reviewers are then highly likely to be together in a collusive group.

\noindent \paragraph{Product tightness(PT).} It is the ratio of the number of common products to the number of products reviewed by all the members in $g$ \cite{Song}.
\begin{equation}
    PT(g) =\frac{|\bigcap_{i \in  R(g)} P_i|}{|\bigcup_{i \in R(g)} P_i|}L(g)
\end{equation}
Group members reviewing
certain number of products and not reviewing any other products are more likely to indulge in fraud activities.

\noindent \paragraph{Rating variance(RV).} Group spammers tend to give similar ratings while reviewing any product. Let \(S^2(p, g)\) be the variance
of the rating scores of product $p$ by reviewers in $g$. We
take the average variance for all target products. The variance degree $[0.5, 1)$ is converted into spam score between $(0, 1]$.
\begin{equation}
RV(g) =2L(g)\Bigg(1- \frac{1}{1+e^{-  \frac{\sum_{p\in P(g)}  S^2(p,g)}{|P(g)|}}} \Bigg) 
\end{equation}

\noindent \paragraph{Product reviewer ratio(RR).} It is defined as the maximum value of the ratio
of the number of reviewers in $R(g)$ who reviewed product $p\in P(g)$ 
to the number of all the reviewers of $p$, denoted by $|Rev(p)|$:
\begin{equation}
RR(g) = \max_{p \in P(g)}  \frac{ \sum_{i\in R(g)} \{1: p\in P_i\}}{|Rev(p)|}
\end{equation}
If a product is mainly reviewed by the reviewers in $g$, then the group is highly likely to be a spammer group.

\noindent \paragraph{Time window(TW).} Fraudsters in a group are likely
to post fake reviews during a short-time interval. Given a group $g$, and a product $p \in P_g$, we define the time-window based spamicity as 

\[
    TW(g,p)= 
\begin{cases}
    1-\frac{SD_{p}}{T},& \text{if } SD_{p}<=T\\
    0,              & \text SD_{p}>T
\end{cases}
\]
\begin{equation}
    TW(g) = \frac{\sum_{p\epsilon Pg} TW(g,p)}{|P(g)|}\cdot L(g),
\end{equation}
where $SD_p$ is the standard deviation of review time for a product p reviewed by reviewers in group $g$. $T$ is a time threshold (set to 30 days in our experiments as suggested in \cite{Wang2018}). 

\begin{algorithm}[tb]
\caption{ExtractGroups}\label{algo:eg}
\begin{algorithmic}[1]
\algnotext{EndFor}
\algnotext{EndIf}
\State $\textbf{Initialize:}$
\State $\hskip 0.3cm  {CCgroups} \gets \text{set()}$ \Comment{Set of candidate groups}
\State $\hskip 0.3cm  {G(V,E)} \gets \text{Edge attributed graph}$
\State $\hskip 0.3cm  {CSet} \gets \text{\{\}}$ \Comment{Potential merged groups}
\State $G,CCgroups\gets\groupdetector(G,CCgroups)$
\State $\textbf{Iterate:}$
\State $\hskip 0.3cm  {G'(V',E')} \gets \text{Attributed line graph of G}$
\State $\hskip 0.3cm G', CCgroups=\groupdetector(G',CCgroups)$
\State $\hskip 0.3cm  {G} \gets {G'}$
\State $\textbf{until }|E| > 1$
\State $\textbf{return }{ CCgroups}$
\Function{\groupdetector}{$G,CCgroups$}\label{line:groupdec}
\For{each isolated node $v_i$ in $G$} \label{egline13}
\State $CCgroups.add(v_i)$ and remove $v_i$ from $G$ \label{egline15}
\EndFor
\For{each pair $(e_i,e_j)\in E\times E$} \label{egline16}
\If{$a_i^e \subset a_j^e$}
\If{$\frac{\bigcap_{m\in a_j^e} P_m}{\bigcup_{m\in a_j^e} P_m} >0.5$}
\State $CSet(a_i^e)=CSet(a_i^e) \cup a_j^e$
\Else
\If{$\frac{\bigcap_{m\in \{a_j^e\setminus a_i^e\}} P_m}{\bigcup_{m\in \{a_j^e\setminus a_i^e\}} P_m} >0.5$}
\State $CCgroups.add (a_j^e\setminus a_i^e)$
\EndIf
\EndIf
\EndIf
\EndFor
\For{each $e_i \in E$} 
\State $k$=$CSet(a_i^e)$
\State $\text{$CCgroups$.add($k$) and remove $k$ from $G$}$  \label{egline26}
\EndFor
\For{each connected component $c$  with $|c|>2$}\label{egline27}
\State $CCgroups.add(c)$ and remove $c$ from $G$ \label{egline29}
\EndFor
\For{each group $g \in CCgroups$ }\label{egline30}
\If {$CollectiveScore(g)<=\tau_{spam}$}
\State $CCgroups.remove(g)$\label{egline32}
\EndIf
\EndFor
\State \textbf{return }$G,CCgroups$
\EndFunction
\end{algorithmic}
\end{algorithm}

\subsection{Detection of Candidate Fraud Groups}\label{sec:groupdetection}
We propose a novel graph-based candidate group detection method based on  the ``coherence principle''.

\begin{hypothesis}
 [\textbf{Coherence Principle}] Fraudsters in a group are coherent in terms of -- (a) the products they review, (b) ratings they give to the products, and (c) the time of reviewing the products. 
\end{hypothesis}

We show that each component of this hypothesis is statistically significant (see Sec. \ref{sec:result}). 
We incorporate these three factors into a novel {\bf attributed product-product graph} construction --  $G(V,E)$ such that each $v_{ij}\in V$ indicates a product-rating pair $(p_i,r_j)$ and its attribute   $a_{ij}^{v}$ consists of the set of reviewers who rated $p_i$ with $r_j$. An edge $e_{(ij,mn)}=\langle u_{ij},v_{mn} \rangle \in E$ indicates the co-reviewing and co-rating patterns of two products $p_i$ and $p_m$ with rating $r_j$ and $r_n$, respectively. The edge attribute $a_{(ij,mn)}^{e}$ indicates  the set of co-reviewers  $R_{(ij,mn)}$  who reviewed both $p_i$ and $p_m$ and gave same ratings $r_j$ and $r_n$ respectively within the same time $\tau_t$ (defined in Sec. \ref{sec:ranking}).   Note that  edge $e_{(ij,in)}$ connecting same product with different ratings wont exist in $G$ as we assume that a reviewer is not allowed to give multiple reviews/ratings to a single product.       

 We then propose {\tt ExtractGroups} (pseudecode in Algo~\ref{algo:eg}, a toy example in Supplementary \citep{si}),
 a novel group detection algorithm that takes $G$ as an input and executes a series of operations through \groupdetector() (Line \ref{line:groupdec}) -- isolated nodes are first removed (Lines~\ref{egline13}-\ref{egline15}), edge attributes are then merged and removed if Jaccard similarity (JS) of product sets that the corresponding reviewers reviewed is greater than a threshold (set as $0.5$). Any group of reviewers eliminated due to the consideration of only common reviewers during the creation of edges is also checked through JS in order to avoid loosing any potential candidate groups (Lines~\ref{egline16}-\ref{egline26}). Before proceeding to the next iteration, connected components containing more than two nodes are removed (Lines \ref{egline27}-\ref{egline29}). We define $CollectiveScore$ as the average of six group level indicators defined in Sec. \ref{sec:indicator}, and consider those as potential groups whose $CollectiveScore$ exceeds $\tau_{spam}$ (Lines~\ref{egline30}-\ref{egline32}, defined in Sec. \ref{sec:metric}).

 It then converts the remaining structure of $G$ into an attributed line graph (edges converted into vertices and vice versa) $G'(V',E')$ as follows: $v'_{(ij,mn)}\in V'$ corresponds to $e_{(ij,mn)}$ in $G$ and $a^{v'}_{(ij,mn)}=a_{(ij,mn)}^{e}$; an edge $e'_{(ij,mn,ab)}=\langle v'_{(ij,mn)}, v'_{(ij,ab)}  \rangle$ represents co-reviewing and co-rating patterns of products $p_i$, $p_m$ and $p_a$; the corresponding edge attribute is $a_{(ij,mn,ab)}=a_{(ij,mn)}^{e}\cap a_{(ij,ab)}^{e}$.
 $G'$ is again fed into \groupdetector\ in the next iteration. Essentially, in each iteration, we keep clubbing reviewers together, who exhibit coherent reviewing patterns. The iteration continues until none of the edges remain in the resultant graph, and a set of candidate fraud groups are returned.

The worst case {\bf time complexity} of {\tt ExtractGroups} is  $\mathcal{O}(k|E|^2)$, where $k$ is the number of iterations, and $|E|$ is the number of edges in $G$ (see Supplementary \citep{si} for more details).

\begin{theorem}[{\bf Theorem of convergence}]
\label{theorem1}
{\tt ExtractGroups} will converge within a finite number of iterations.
\end{theorem}
See Supplementary \citep{si} for the proof.

\subsection{Ranking of Candidate Fraud Groups}\label{sec:ranking}
Once candidate fraud groups are detected, \system\ ranks these groups based on their spamicity. It involves two steps -- mapping reviewers into an embedding space based on their co-reviewing patterns, and  ranking groups based on how close the constituent reviewers are in the embedding space.  

\subsubsection{\reviewervec: Embedding Reviewers}
Our proposed embedding method, \reviewervec\ is motivated by \citep{Wang2018}. Given two reviewers $i$ and $j$ co-reviewing a product
$p \in P$ by writing reviews $c_p^i$ and $c_p^j$   with the rating $r_p^i$ and $r_p^j$ at time $t_p^i$ and $t_p^j$ respectively, we define the {\em collusive spamicity} of $i$, $j$ w.r.t. $p$ as:
\begin{equation}\small
    Coll(i,j,p)=
\begin{cases}
0, & |t^p_{i}-t^p_{j}|>\tau_t \vee	 |r^p_{i}-r^p_{j}|\geq \tau_r \\
\zeta(i,j,p), &  \text{otherwise}
\end{cases}
\end{equation}
where, 
\begin{equation*}\small
\begin{split}
\zeta(i,j,p)=& s^p\Bigg[\alpha \Bigg(1-\frac{|t_p^{i}-t_p^{j}|}{\tau_t}\Bigg)+\beta \Bigg(1-\frac{|r_p^{i}-r_p^{j}|}{\tau_r}\Bigg) \\
&+ \gamma. Cosine(c_p^{i},c_p^{j})\Bigg]
\end{split}
\end{equation*}
Where $s^p=\frac{2}{1+e^{-(\max\limits_{q\in P}Rev(q)-Rev(p))^\theta+2^\theta}}-1$. Coefficients $\alpha$, $\beta$ and $\gamma$ control the importance of time, rating and the similarity of review content, respectively ($0\leq\alpha,\beta,\gamma\leq1$ and $\alpha+\beta+\gamma=1$). We set $\gamma>\alpha,\beta$  as same review content signifies more collusion as compared to the coherence in ratings and time \citep{Wang2018}. 

$s^p$ is the degree of suspicion of product $p$. $\tau_t$, $\tau_r$ and $\theta$ are the parameters of the model. If the posting time difference among reviewers $i$ and $j$ on $p$ is beyond $\tau_t$  or their rating on $p$ deviates beyond $\tau_r$ (where $\tau_r=(\max-\min)x\%$, where $\max$ and $\min$ are the maximum and minimum ratings that a product can achieve respectively), we do not consider this co-reviewing pattern. {\tt ExtractGroups} achieves best results with $\tau_t=20$ and $\tau_r=(\max-\min)20\%$ (see Sec. \ref{sec:evaluateranking}). 

$\zeta(i,j,p)$ is the collusiveness between two reviewers w.r.t. product $p$ they co-reviewed; however there might be other reviewers who reviewed $p$ as well. Lesser the number of reviewers of $p$, more is the probability that $i$ and $j$ colluded. This factor is handled by $s^p$ after passing it through a sigmoid function. $\theta$ is a normalizing factor which ranges between $[0,1]$ (set as $0.4$ \citep{Wang2018}). We take the cosine similarly of two reviews  $c_p^i$ and $c_p^j$ after mapping them into embedding space using Word2Vec \citep{Mikolov:2013}. 

Combining the collusive spamicity of a pair of reviewers across all the products they co-reviewed, we obtain the overall collusive spamicity between two reviewers:
\begin{equation*}\small
\begin{split}
 \Phi(i,j)&=\frac{2}{1+e^{-\sigma(i,j)}}-1\\
\text{where } \sigma(i,j)&=\Bigg[\sum_{p \in \{P_i \cap P_j\}} Coll(i,j,p)\Bigg]\frac{|P_i \cap P_j|}{|P_i \cup P_j|}
\end{split}
\end{equation*}

We then create a reviewer-reviewer spammer graph \citep{Wang2018} which is a bi-connected and weighted graph  $G_s = (V_s, E_s)$, where  $V_s$ corresponds to the set of reviewers $R$, and two reviewers $i$ and $j$ are connected by an edge $e_{ij}^s\in E_s$ with the weight  $W_{ij}^s=\Phi(i,j)$. 
Once $G_s$ is created, we use state-of-the-art node embedding method to generate node (reviewer) embeddings (see Sec. \ref{sec:evaluateranking}).

\subsubsection{Ranking Groups}
For each detected group, we calculate the density of the group based on (Euclidean) distance of each reviewer in the group with the group's centroid in the embedding space. An average of all distances is taken as a measure of spamicity of the group. Let $\vec{i}$ be the vector representation of reviewer $i$ in the embedding space. The group spam score $Spam(g)$ of  group $g$ is measured as:
\begin{equation}\label{eq:spam}\small
    Spam(g)=\frac{1}{|R(g)|} \sum_{i\in R(g)} \Bigg[\vec{i} - \big[\frac{1}{|R(g)|} \sum_{i\in R(g)} \vec{i}\big]\Bigg]^2
\end{equation}

\section{Datasets}
We collected four real-world datasets  -- {\bf YelpNYC}: hotel/restaurant reviews of New York city \cite{Shebuti_SIGKDD}; {\bf YelpZip}: aggregation of reviews on restaurants/hotels from a number of areas with continuous zip codes starting from New York city \cite{Shebuti_SIGKDD};  {\bf Amazon}: reviews on musical instruments
\cite{He:2016:UDM:2872427.2883037}, and {\bf Playstore}: reviews of different applications available on Google Playstore. Fake reviews and spammers are already marked in both the Yelp datasets ~\cite{Shebuti_SIGKDD}. For the remaining datasets, we employed three human experts\footnote{They were social media experts, and their age ranged between 25-35.} to label spammers based on the instructions mentioned in \cite{Shojaee:2015,Arjun_ACM}\footnote{\url{https://www.marketwatch.com/story/10-secrets-to-uncovering-which-online-reviews-are-fake-2018-09-21}}.
% They were also given full liberty to apply their own experience. 
The inter-rater agreement was $0.81$ and $0.79$ (Fleiss’ multi-rater kappa) for  Amazon and Playstore, respectively.  
Table \ref{tab:dataset} shows the statistics of the datasets.

\begin{table}[]
 \centering
      \begin{tabular}{r|r|r|r}
        \toprule Dataset & \# Reviews & \# Reviewers & \# Products \\ \midrule
        YelpNYC & 359052&160225&923 \\ 
        YelpZIP & 608598&260227&5044\\ 
        Amazon & 10261&1429&900\\ 
        Playstore & 325424&321436&192\\ \bottomrule
      \end{tabular}
      \caption{Statistics of four datasets.}\label{tab:dataset}
\end{table}

\begin{table*}[]
    \centering
    \scalebox{0.82}{
    \begin{tabular}{l|b|rr|a|b|rr|a|b|rr|a|b|rr|a}
    \toprule
       \multirow{2}{*}{Method} & \multicolumn{4}{c|}{YelpNYC} & \multicolumn{4}{c|}{YelpZIP} & \multicolumn{4}{c|}{Amazon} & \multicolumn{4}{c}{Playstore}
       \\\cline{2-17} 
        & $|G|$ & $GS$ & $RCS$ & ND & $|G|$ & $GS$ & $RCS$ &ND & $|G|$ & $GS$ & $RCS$ &ND &$|G|$&  $GS$ & $RCS$ & ND \\\midrule

      GGSpam &1218&0.574&0.218& 0.567 &1167&0.629&0.219& 0.563 &144&0.131&0.250& 0.230 &1213& 0.749 & 0.010 &0.464\\ 
      GSBP &809&0.562&0.173&0.521& 807&0.478&0.265 &0.520&115&0.416&0.260 &0.689& 250&0.744&0.016&0.474 \\
      GSRank &998&0.102&0.313& 0.569  &1223 & 0.132 & 0.054 & {\bf 0.706} & 2922  & 0.293 & 0.309 & 0.144 & 994 & 0.577 & 0.018 & 0.476 \\\midrule
        \system\textsubscript{R}&4399&0.124&0.021 &--- & 6815&0.139&0.031 & --- & 197&0.234&0.290 &--- & 385& 0.372& 0.005& ---\\
      \system\textsubscript{T} &152&0.237&0.069&--- &3666&0.648&0.271 &--- & 807  &0.698&0.207 &--- & 200 & 0.458 & 0.007 & ---\\\midrule
      \system &1118&\textbf{0.731}&\textbf{0.348}  & {\bf 0.603} & 4574&\textbf{0.667}&\textbf{0.287} & 0.602& 713&\textbf{0.718}&\textbf{0.314} & {\bf 0.768} & 940 &{\bf 0.841} &  {\bf 0.018} & {\bf 0.789}\\\bottomrule
    \end{tabular}}
    % \vspace{-3mm}
    \caption{Performance  of the competing methods:  GGSpam \cite{Wang2018}, GSBP \cite{Song}, GSRank \cite{Arjun_www}, \system\textsubscript{R}, \system\textsubscript{T}, and \system. Number of groups  detected ($|G|$) are mentioned after removing groups of size less than 2 (cyan regions). Accuracy for the group detection (white regions) and ranking (gray  regions) is reported in terms of EMD (the higher, the better), and NDCG@50 (ND) respectively. \system\textsubscript{R} and \system\textsubscript{T} are used only for group detection. The ranking methods of all baselines are run on the groups detection by \system.}
    \label{tab:accuracy}
    % \vspace{-5mm}
\end{table*}

\section{Experimental Results}\label{sec:result}
\paragraph{Statistical validation.} To measure the statistical significance of each component (say, products reviewed by the group members) of Hypothesis 1 (Sec. \ref{sec:groupdetection}), we randomly generate pairs of reviewers (irrespective of the groups they belong to) and measure how their co-reviewing patterns (cumulative distribution) are different from the case if a pair of reviewers co-occur together in the same group. We hypothesize that both these patterns are different (whereas null hypothesis rejects our hypothesis). We observe that the difference is statistically significant as $p<0.01$ (Kolmogorov-Smirnov test). See Supplementary \cite{si} for more discussion. We then perform evaluation in two stages -- quality of the detected  groups, and quality of the ranking algorithms.
\subsection{Baseline Methods} We consider  three existing methods (see Sec. \ref{sec:relatedwork} for their details) as baselines for both the evaluations -- (i) {\bf GSBP} \cite{Song},(ii) {\bf GGSpam} \cite{Wang2018},and (iii) {\bf GSRank} \cite{Arjun_www}.

\subsection{Evaluating Candidate Groups}\label{sec:groupeval}
Along with the three baselines mentioned above, we also consider two variations of \system\ as baselines for group detection: {\bf \system\textsubscript{R}} constructs the attributed product-product graph $G$ based only on co-rating without considering the time of reviewing; and {\bf \system\textsubscript{T}} constructs $G$ based only on same co-reviewing time without considering co-rating. This also helps in justifying why both time and rating are important in constructing $G$ for group detection. \cite{Arjun_www} suggested to use cumulative distribution (CDF) of {\em group size} (GS) and {\em review content similarity} (RCS) to evaluate the quality of the spam groups. Here we discard groups with size less than $2$. {\bf Group Size (GS)} favors large fraud groups as large groups are more damaging than smaller ones: $GS(g)=\frac{1}{1+e^{\textbf{-}(|R(g)|-2)}}$. {\bf Review Content Similarity (RCS)} captures inter-reviewer review content similarity (as spammers copy reviews of each other): $RCS(g)=\max_{p\in P(g)}\{\frac{1}{|R(g)|^2} \sum_{(i,j)\in R(g)\times R(g)} cosine(c_p^{i},c_p^{j})\}$.

The larger the deviation of each distribution from the vertical axis (measured in terms of Earth Mover's Distance (EMD)), the better the quality of the detected method \cite{Arjun_www}.

\begin{figure}
\centering
\centering
    \includegraphics[width=\columnwidth]{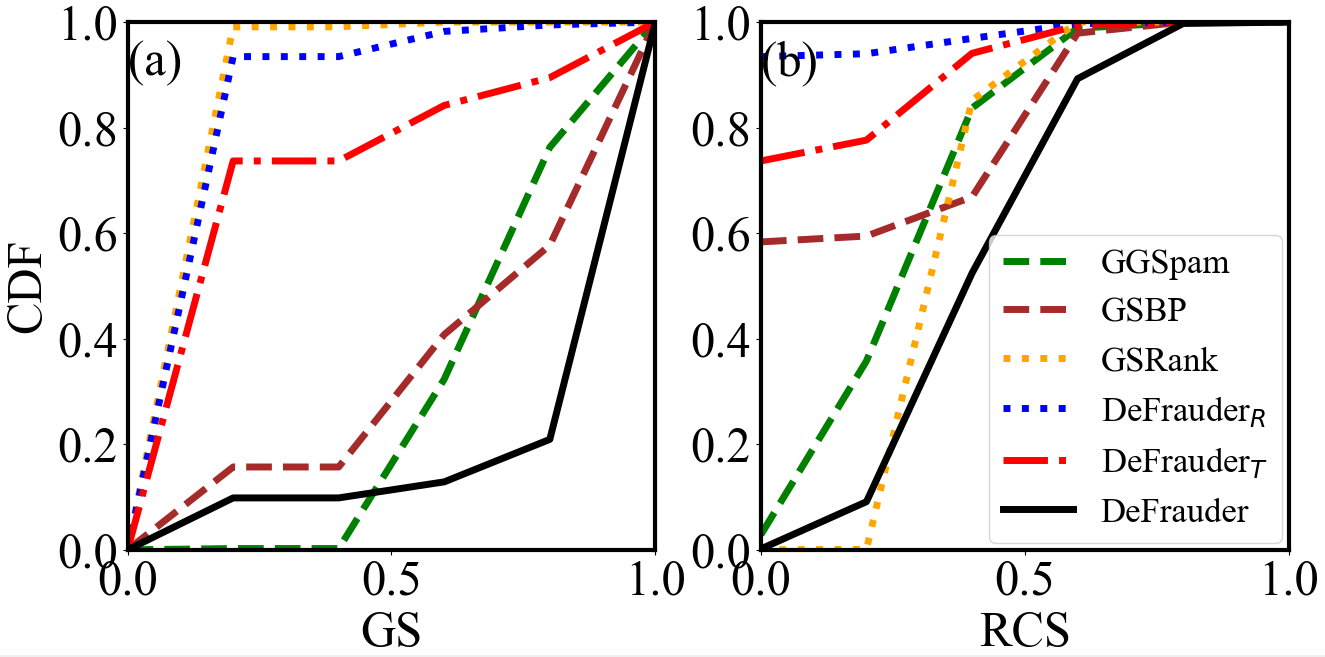} 
\caption{CDF of (a) GS and (b) RCS for YelpNYC. The more the distance of a CDF (corresponding to a method) from y-axis, the better the performance of the method. }\label{fig:CDF}
\end{figure}

\paragraph{Comparison.}\label{sec:metric} We choose the following parameter values as default based on our parameter selection strategy (Fig. \ref{fig:parameter}): $\tau_t$: 20 days, $\tau_{spam}=0.4$.
The number of groups we obtain from different datasets is reported in Table \ref{tab:accuracy}.  Fig. \ref{fig:CDF} shows the CDF of GS and RCS for all the competing methods on YelpNYC\footnote{Results are similar on the other datasets, and thus omitted.}, which is further summarized  quantitatively (in terms of EMD) in Table \ref{tab:accuracy}.  \system\ outperforms the best baseline by  11.35\% and 4.67\% higher relative EMD  (averaged over four datasets) for GS and RCS, respectively. We also notice that \system\textsubscript{T} performs better than \system\textsubscript{R}, indicating that temporal coherence is more important than rating coherence in detecting potential groups.

\subsection{Evaluating Ranking Algorithm}\label{sec:evaluateranking}
We use \textbf{NDCG@k}, a standard graded evaluation metric.  Since each reviewer was labeled as fraud/genuine, we consider the graded relevance value (ground-truth relevance score used in Normalized Discounted Cumulative Gain (NDCG) \cite{manning2010introduction}) for each group as the fraction of reviewers in a group, who are marked as fraud. The candidate groups are then ranked by each competing method, and  top $k$ groups are judged based on NDCG.

\begin{figure}
\centering
\scalebox{0.84}{
\includegraphics[width=\columnwidth]{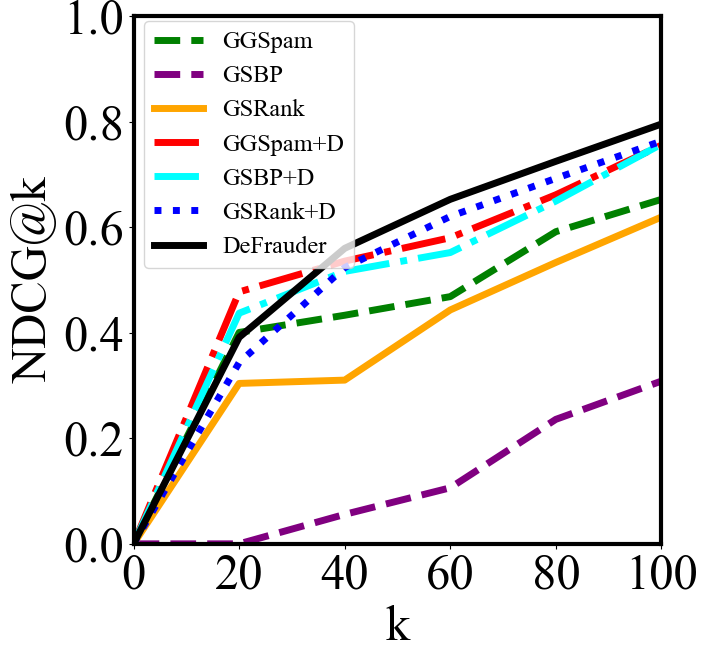}}
\caption{Performance on YelpNYC.
Baselines are run with their detected groups as well as with the groups (+D) detected by \system\ (same naming convention as in Table \ref{tab:accuracy}).}\label{fig:ndcg}
\end{figure}

\begin{figure}
\begin{subfigure}{\columnwidth}
    \centering
    \scalebox{0.75}{
    \includegraphics[width=\columnwidth]{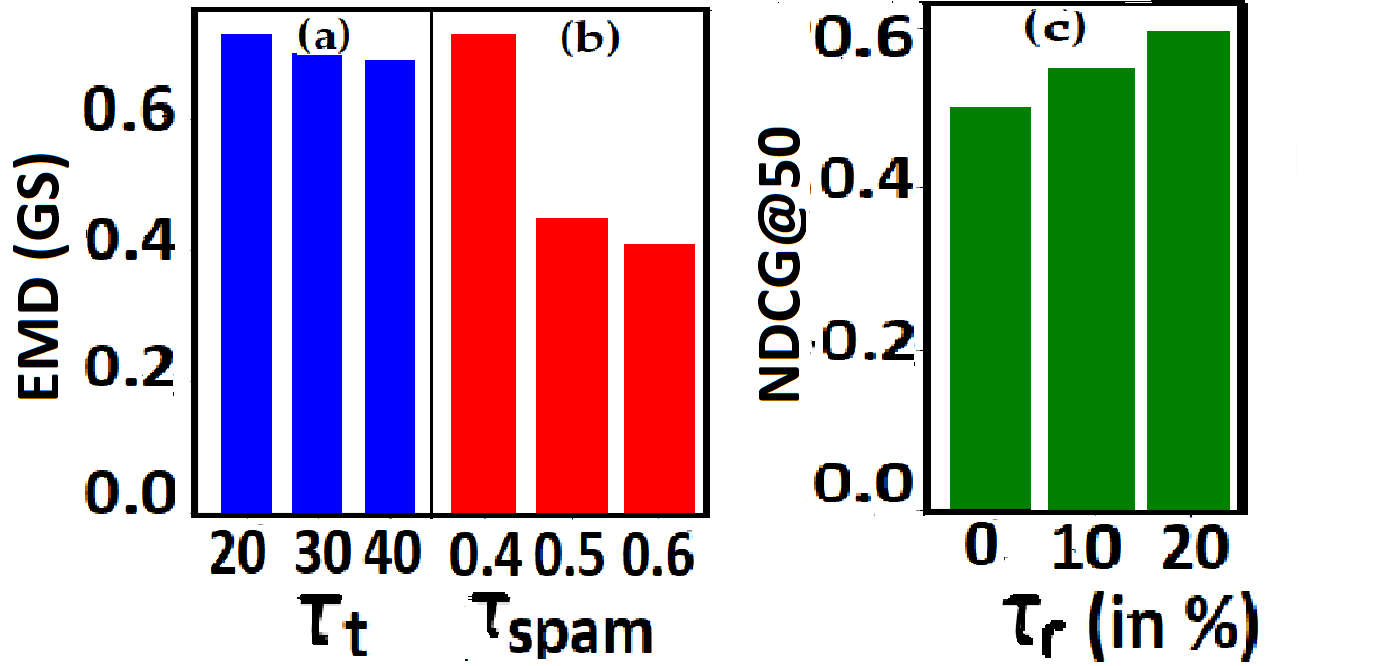}}
\end{subfigure}
\hfill
\begin{subfigure}{\columnwidth}
    \centering
    \scalebox{0.6}{
    \includegraphics[width=\columnwidth]{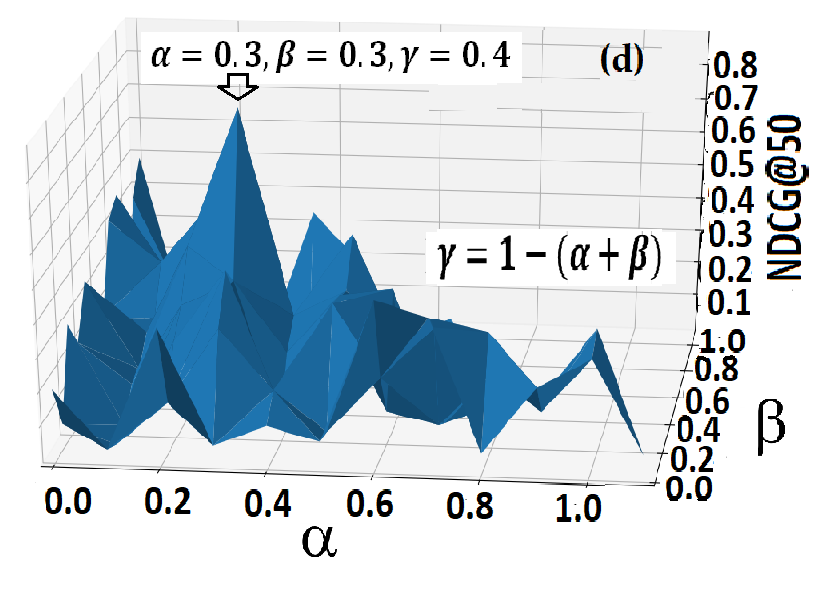}} 
\end{subfigure}
\hfill
    % \includegraphics[width=\columnwidth]{PLOTS/YELPNYC/Parameter.png}
    % \vspace{-3mm}
\caption{Parameter selection on YelpNYC. We vary each parameter keeping others as default. Since $\tau_t=20$ produces best results for group detection, we kept this value for ranking as well. }\label{fig:parameter}
% \vspace{-5mm}
\end{figure}

\paragraph{Comparison.} We choose the following parameter values as default based on our parameter selection strategy (Fig. \ref{fig:parameter}): $\alpha=\beta=0.3$, $\gamma=0.4$, $\tau_t= 20$ days, $\tau_r=(\max- \min)20\%$ $\tau_{spam}=0.4$. We use Node2Vec \cite{Grover:2016} for embedding\footnote{We tried with DeepWalk \cite{Perozzi:2014} and LINE \cite{Tang:2015} and obtained worse results compared to Node2Vec.}.
Fig. \ref{fig:ndcg} shows the performance of the competing methods for different values of $k$ (top $k$ groups returned by each method).  Since \system\ produces better groups (Sec. \ref{sec:groupeval}), we also check how the ranking method of each baseline performs on the groups detected of \system. Fig. \ref{fig:ndcg} shows  that with \system's group structure, GGSpam and GSBP show better performance than \system\ till $k=40$; after that \system\ dominates others. However, all the baselines perform poor with their own detected groups.  This result also indicates the efficiency of our group detection method. Table \ref{tab:accuracy} reports that \system\ beats other methods across all the datasets except YelpZIP on which GSRank performs better with \system's detected groups. Interestingly, no single baseline turns out to be the best baseline across datasets. 
Nevertheless,  \system\ outperforms the best baseline (varies across datasets) by 17.11\% higher relative NDCG@50 (averaged over all the datasets).   

%------------------------------------------
\section{Conclusion}
In this paper, we studied the problem of fraud reviewer group detection in customer reviews. We established the principle of coherence among fraud reviewers in a group in terms of their co-reviewing patterns.  This paper contributed in four directions: {\bf Datasets:} We collected and annotated two new datasets which would be useful to the cybersecurity community; {\bf Characterization:} We explored several group-level behavioral traits to model inter-personal collusive dynamics in a group; {\bf Method:} We proposed \system, a novel method to detect and rank fraud reviewer groups; {\bf Evaluation:} Exhaustive experiments were performed on four datasets to show the superiority of \system\ compared to five baselines.

\section*{Acknowledgements} The work was partially funded by Google Pvt. Ltd., DST (ECR/2017/00l691) and Ramanujan Fellowship. The authors also acknowledge the support of the Infosys Centre of AI, IIIT-Delhi, India.

\newpage
% {\small 
\bibliographystyle{named}

\begin{thebibliography}{}

\bibitem[\protect\citeauthoryear{Akoglu \bgroup \em et al.\egroup
 }{2013}]{Akoglu_ICWSM}
Leman Akoglu, Rishi Chandy, and Christos Faloutsos.
\newblock Opinion fraud detection in online reviews by network effects.
\newblock In {\em ICWSM}, pages 985–--994, 2013.

\bibitem[\protect\citeauthoryear{Allahbakhsh \bgroup \em et al.\egroup
  }{2013}]{allahbakhsh2013collusion}
Mohammad Allahbakhsh, Aleksandar Ignjatovic, Boualem Benatallah, Elisa Bertino,
  Norman Foo, et~al.
\newblock Collusion detection in online rating systems.
\newblock In {\em Asia-Pacific Web Conference}, pages 196--207. Springer, 2013.

\bibitem[\protect\citeauthoryear{Chetan \bgroup \em et al.\egroup
  }{2019}]{Chetan:2019}
Aditya Chetan, Brihi Joshi, Hridoy~Sankar Dutta, and Tanmoy Chakraborty.
\newblock Corerank: Ranking to detect users involved in blackmarket-based
  collusive retweeting activities.
\newblock In {\em WSDM}, pages 330--338, 2019.

\bibitem[\protect\citeauthoryear{Dhawan \bgroup \em et al.\egroup }{}]{si}
Sarthika Dhawan, Siva Charan~Reddy Gangireddy, Shiv Kumar, and Tanmoy
  Chakraborty.
\newblock Defrauder: Supplementary.
\newblock https://github.com/LCS2-IIITD/DeFrauder, 2019.

\bibitem[\protect\citeauthoryear{Dutta \bgroup \em et al.\egroup
  }{2018}]{dutta2018retweet}
Hridoy~Sankar Dutta, Aditya Chetan, Brihi Joshi, and Tanmoy Chakraborty.
\newblock Retweet us, we will retweet you: Spotting collusive retweeters
  involved in blackmarket services.
\newblock In {\em ASONAM}, pages 242--249, 2018.

\bibitem[\protect\citeauthoryear{Fei \bgroup \em et al.\egroup
  }{2013}]{Fei_AAAI}
Geli Fei, Arjun Mukherjee, Bing Liu, Meichun Hsu, Malu Castellanos, and
  Riddhiman Ghosh.
\newblock Exploiting burstiness in reviews for review spammer detection.
\newblock In {\em ICWSM}, 2013.

\bibitem[\protect\citeauthoryear{Grover and Leskovec}{2016}]{Grover:2016}
Aditya Grover and Jure Leskovec.
\newblock Node2vec: Scalable feature learning for networks.
\newblock In {\em SIGKDD}, pages 855--864, 2016.

\bibitem[\protect\citeauthoryear{He and
  McAuley}{2016}]{He:2016:UDM:2872427.2883037}
Ruining He and Julian McAuley.
\newblock Ups and downs: Modeling the visual evolution of fashion trends with
  one-class collaborative filtering.
\newblock In {\em WWW}, pages 507--517, 2016.

\bibitem[\protect\citeauthoryear{Hill}{2018}]{WinNT}
Catey Hill.
\newblock 10 secrets to uncovering which online reviews are fake.
\newblock
  https://www.marketwatch.com/story/10-secrets-to-uncovering-which-online-reviews-are-fake-2018-09-21,
  2018.

\bibitem[\protect\citeauthoryear{Kou \bgroup \em et al.\egroup
  }{2004}]{kou2004survey}
Yufeng Kou, Chang-Tien Lu, Sirirat Sirwongwattana, and Yo-Ping Huang.
\newblock Survey of fraud detection techniques.
\newblock In {\em ICNSC}, pages 749--754, 2004.

\bibitem[\protect\citeauthoryear{Lim \bgroup \em et al.\egroup
  }{2010}]{Lim_CIKM10}
Ee-Peng Lim, Viet-An Nguyen, Nitin Jindal, Bing Liu, and Hady~Wirawan Lauw.
\newblock Detecting product review spammers using rating behaviors.
\newblock In {\em CIKM}, pages 939--948. ACM, 2010.

\bibitem[\protect\citeauthoryear{Manning \bgroup \em et al.\egroup
  }{2010}]{manning2010introduction}
Christopher Manning, Prabhakar Raghavan, and Hinrich Sch{\"u}tze.
\newblock Introduction to information retrieval.
\newblock {\em Natural Language Engineering}, 16(1):100--103, 2010.

\bibitem[\protect\citeauthoryear{Mikolov \bgroup \em et al.\egroup
  }{2013}]{Mikolov:2013}
Tomas Mikolov, Ilya Sutskever, Kai Chen, Greg Corrado, and Jeffrey Dean.
\newblock Distributed representations of words and phrases and their
  compositionality.
\newblock In {\em NIPS}, pages 3111--3119, 2013.

\bibitem[\protect\citeauthoryear{Mukherjee \bgroup \em et al.\egroup
  }{2012}]{Arjun_www}
Arjun Mukherjee, Bing Liu, and Natalie Glance.
\newblock Spotting fake reviewer groups in consumer reviews.
\newblock In {\em WWW}, pages 191–--200, New York, April 2012.

\bibitem[\protect\citeauthoryear{Mukherjee \bgroup \em et al.\egroup
  }{2013a}]{Arjun_ACM}
Arjun Mukherjee, Abhinav Kumar, Junhui Liu, Bing~Wang, Meichun Hsu, Malu
  Castellanos, and Riddhiman Ghosh.
\newblock Spotting opinion spammers using behavioral footprints.
\newblock In {\em SIGKDD}, Chicago,USA, August 2013.

\bibitem[\protect\citeauthoryear{Mukherjee \bgroup \em et al.\egroup
  }{2013b}]{mukherjee2013yelp}
Arjun Mukherjee, Vivek Venkataraman, Bing Liu, and Natalie Glance.
\newblock What yelp fake review filter might be doing?
\newblock In {\em ICWSM}, pages 1--12, 2013.

\bibitem[\protect\citeauthoryear{Ott \bgroup \em et al.\egroup
  }{2011}]{ott2011finding}
Myle Ott, Yejin Choi, Claire Cardie, and Jeffrey~T Hancock.
\newblock Finding deceptive opinion spam by any stretch of the imagination.
\newblock In {\em ACL-HLT}, pages 309--319, 2011.

\bibitem[\protect\citeauthoryear{Perozzi \bgroup \em et al.\egroup
  }{2014}]{Perozzi:2014}
Bryan Perozzi, Rami Al-Rfou, and Steven Skiena.
\newblock Deepwalk: Online learning of social representations.
\newblock In {\em SIGKDD}, pages 701--710, 2014.

\bibitem[\protect\citeauthoryear{Rayana and Akoglu}{2015}]{Shebuti_SIGKDD}
Shebuti Rayana and Leman Akoglu.
\newblock Collective opinion spam detection: Bridging review networks and
  metadata.
\newblock In {\em SIGKDD}, pages 985–--994, Sydney, NSW, Australia, August
  2015.

\bibitem[\protect\citeauthoryear{Shojaee \bgroup \em et al.\egroup
  }{2015}]{Shojaee:2015}
Somayeh Shojaee, Azreen Azman, Masrah Murad, Nurfadhlina Sharef, and Nasir
  Sulaiman.
\newblock A framework for fake review annotation.
\newblock In {\em UKSIM-AMSS}, pages 153--158, 2015.

\bibitem[\protect\citeauthoryear{Tang \bgroup \em et al.\egroup
  }{2015}]{Tang:2015}
Jian Tang, Meng Qu, Mingzhe Wang, Ming Zhang, Jun Yan, and Qiaozhu Mei.
\newblock Line: Large-scale information network embedding.
\newblock In {\em WWW}, pages 1067--1077, 2015.

\bibitem[\protect\citeauthoryear{Wang \bgroup \em et al.\egroup
  }{2011}]{Wang_ICDM}
Guan Wang, Sihong Xie, Bing Liu, and S~Yu Philip.
\newblock Review graph based online store review spammer detection.
\newblock In {\em ICDM}, pages 1242--1247. IEEE, 2011.

\bibitem[\protect\citeauthoryear{Wang \bgroup \em et al.\egroup
  }{2012}]{Guan_ACM}
Guan Wang, Sihong Xie, Bing Liu, and Philip~S Yu.
\newblock Identify online store review spammers via social review graph.
\newblock {\em ACM TIST}, 3(4):61:1--61:21, 2012.

\bibitem[\protect\citeauthoryear{Wang \bgroup \em et al.\egroup }{2016}]{Song}
Zhuo Wang, Tingting Hou, Dawei Song, Zhun Li, and Tianqi Kong.
\newblock Detecting review spammer groups via bipartite graph projection.
\newblock {\em The Computer Journal}, 59(6):861--874, 2016.

\bibitem[\protect\citeauthoryear{Wang \bgroup \em et al.\egroup
  }{2018a}]{wang2018gslda}
Zhuo Wang, Songmin Gu, and Xiaowei Xu.
\newblock Gslda: Lda-based group spamming detection in product reviews.
\newblock {\em Applied Intelligence}, 48(9):3094--3107, 2018.

\bibitem[\protect\citeauthoryear{Wang \bgroup \em et al.\egroup
  }{2018b}]{Wang2018}
Zhuo Wang, Songmin Gu, Xiangnan Zhao, and Xiaowei Xu.
\newblock Graph-based review spammer group detection.
\newblock {\em KIAS}, 55(3):571--597, Jun 2018.

\bibitem[\protect\citeauthoryear{Xu and Zhang}{2015}]{Xu_IEEE}
Chang Xu and Jie Zhang.
\newblock Towards collusive fraud detection in online reviews.
\newblock In {\em ICDM}, pages 1051--1056, 2015.

\bibitem[\protect\citeauthoryear{Xu \bgroup \em et al.\egroup
  }{2013}]{xu2013uncovering}
Chang Xu, Jie Zhang, Kuiyu Chang, and Chong Long.
\newblock Uncovering collusive spammers in chinese review websites.
\newblock In {\em CIKM}, pages 979--988, 2013.

\end{thebibliography}

% }

\end{document}